\documentclass[aps,prl,twocolumn,showpacs]{revtex4}
\usepackage{psfrag}
\usepackage{epsfig}
\usepackage[usenames]{color}
\usepackage{amssymb}
\usepackage{color}

\newcommand     {\beq}[1]         { \begin{equation} #1 \end{equation} }

\begin{document}

\title{Universality behind Basquin's law of fatigue}

\author{F.\ Kun${}^{1,2}\footnote{Electronic 
address:feri@dtp.atomki.hu}$, H.\ A.\ Carmona${}^{2,3}$, J.\ S.\
Andrade Jr.${}^{2,4}$, and H.\ J.\ Herrmann${}^{2}$}

\affiliation{ 
\centerline{${}^1$Department of Theoretical Physics, University of Debrecen, P.\
O.\ Box:5, H-4010 Debrecen, Hungary} \\
\centerline{${}^2$ Computational Physics, IfB, HIF, E12, ETH, H\"onggerberg, 8093
Z\"urich, Switzerland} \\
\centerline{${}^3$ Centro de Ci\^encias e Tecnologia, Universidade Estadual 
do Cear\'a, 60740-903 Fortaleza, Cear\'a, Brazil} \\
\centerline{${}^4$ Departamento de F\'{\i}sica, Universidade Federal
do Cear\'a, 60451-970 Fortaleza, Cear\'a, Brazil}
}

\date{\today}

\begin{abstract}
One of the most important scaling laws of time dependent fracture is
Basquin's law of fatigue, namely, that the lifetime of the
system increases as a power law with decreasing external load
amplitude, $t_f\sim \sigma_0^{-\alpha}$, where the exponent $\alpha$
has a strong material dependence. We show that in spite of the broad
scatter of the Basquin exponent $\alpha$, the fatigue fracture of
heterogeneous materials exhibits intriguing universal features. Based on
stochastic fracture models we propose a generic scaling form for
the macroscopic deformation and show that at the fatigue limit the
system undergoes a continuous phase transition when changing the
external load. On the microlevel, the fatigue fracture proceeds in
bursts characterized by universal power law distributions. We
demonstrate that in a range of systems, including deformation of
asphalt, a realistic model of deformation, and a fiber bundle model,
the system dependent details are contained in Basquin's exponent for
time to failure, and once this is taken into account, remaining
features of failure are universal. 
\end{abstract}


\maketitle

Disordered media subject to sub-critical external loads present a time
dependent macroscopic response and typically fail after a finite time
\cite{zapperi_alava_statmodfrac}. Such time dependent fracture
evidently plays a crucial role in a large variety of physical,
biological, and geological systems, such as the rupture of adhesion
clusters of cells in biomaterials under external stimuli
\cite{schwarz_adhesionbio_prl_2004}, the sub-critical crack growth due
to thermal activation of crack nucleation
\cite{santucci_prl_subcrit_2004,sornette_thermal_prl_2005},
creep \cite{nechad_sornette_prl_2005} and fatigue fracture of materials
\cite{sornette_fusefatigue_prl1992,farkas_crystal_fatigue_prl_2005},
and the emergence of earthquake sequences
\cite{marone_nature_healing_1998}. 
One of the most important scaling laws of time dependent fracture is
the empirical Basquin law of fatigue which states that the
lifetime $t_f$ of samples increases as a power law when the external
load amplitude $\sigma_0$ decreases, $t_f\sim
\sigma_0^{-\alpha}$ \cite{basquin_1910}. The measured values of the
Basquin exponent 
$\alpha$ typically vary over a broad range indicating a strong
dependence on material properties
\cite{basquin_1910,krajcinovic_damagebook_1996,krajcinovic_1}.

In this Letter we study the fatigue fracture of heterogeneous
materials focusing on the underlying microscopic mechanism of the
fatigue process and its relation to the macroscopic time evolution.
We develop two generic models of time dependent fracture, namely, a
fiber bundle model and a discrete element
approach, which both capture the most important
ingredients of the fatigue failure of disordered materials. Analytic
solutions and computer simulations reveal that the models recover the
Basquin law of fatigue, whose exponent is determined by the damage
process.  We show that, as a consequence of healing, a finite fatigue
limit emerges at which the system undergoes a continuous phase
transition from a regime where macroscopic failure occurs at a finite
time to another one exhibiting only partial failure in the system
having an infinite lifetime. Based on analytic solutions, we propose a
generic scaling form for the macroscopic deformation. 
On the microlevel the fatigue of the
material is accompanied by an avalanche activity where bursts of local
breakings are triggered by damage sequences. We demonstrate
analytically that the microscopic bursting activity underlying fatigue
fracture is characterized by universal power law distributions which
implies that the non-universality of the Basquin exponent at the
macro-level is solely due to the specific degradation process of the
material.

First we consider a mean field model of fatigue fracture, namely, a
fiber bundle model (FBM) where fibers fail either due to immediate
breaking or to ageing \cite{kun_asphalt_jstat_2007}. For the load
redistribution after failure events, equal load sharing is assumed so
that all the fibers carry the same load \cite{sornette_prl_78_2140}. 
During the evolution of the system, a fiber breaks instantaneously at
time $t$ when the load on it $p(t)$ exceeds the local tensile strength
$p_{th}^i$ ($i=1, \ldots , N$). All intact fibers accumulate damage
$c(t)$ due to the load $p(t)$ that they have experienced and break
when $c(t)$ exceeds the local damage threshold $c_{th}^i$ ($i=1,
\ldots , N$). The accumulated damage $c(t)$ up to time $t$ is obtained
by integrating over the entire loading history of the specimen $c(t) =
a \int \limits_0^te^{-\frac{(t-t')}{\tau}}p(t')^{\gamma}dt'$, where
$a>0$ is a scale parameter, while the exponent $\gamma>0$ controls the
rate of damage accumulation
\cite{krajcinovic_damagebook_1996,krajcinovic_1}.  To capture damage
recovery in the model due to healing of microcracks
\cite{krajcinovic_damagebook_1996} or thermally
activated rebinding of failed contacts
\cite{schwarz_adhesionbio_prl_2004,marone_nature_healing_1998}, we
introduce a memory term in the above damage law of exponential
form whose characteristic time scale $\tau$ defines the 
memory range of the system
\cite{marone_nature_healing_1998,schwarz_adhesionbio_prl_2004,sornette_thermal_prl_2005}.
Hence, during the time evolution of the bundle, the
damage accumulated over the time interval $t' < (t-\tau)$ heals.
Assuming independence of the two breaking thresholds $p_{th}$ and
$c_{th}$, the macroscopic evolution of the system
under a constant external load $\sigma_0$ can be cast into the form   
\beq{
\sigma_0 = [1-F(c(t))]\left[1-G(p(t))\right]p(t), 
\label{eq:eom}
}
where $G$ and $F$ denote the cumulative distributions of $p_{th}$ and
$c_{th}$, respectively.
We solved Eq.\ (\ref{eq:eom}) analytically obtaining the load $p(t)$ on
the intact fibers at a constant external load $\sigma_0<\sigma_c$,
with the initial condition $p(t=0) = p_0$, where $p_0$ denotes the 
solution of the constitutive equation $\sigma_0 = \left[1-G(p_0)
\right]p_0$ \cite{sornette_prl_78_2140}. Here
$\sigma_c$ denotes the ultimate strength of the
material. The most important input parameters of the model
calculations are $a$, $\gamma$ and $\tau$, which govern the damage
accumulation.

\begin{figure}%[!h]
  \begin{center}
\epsfig{bbllx=20,bblly=465,bburx=350,bbury=750,file=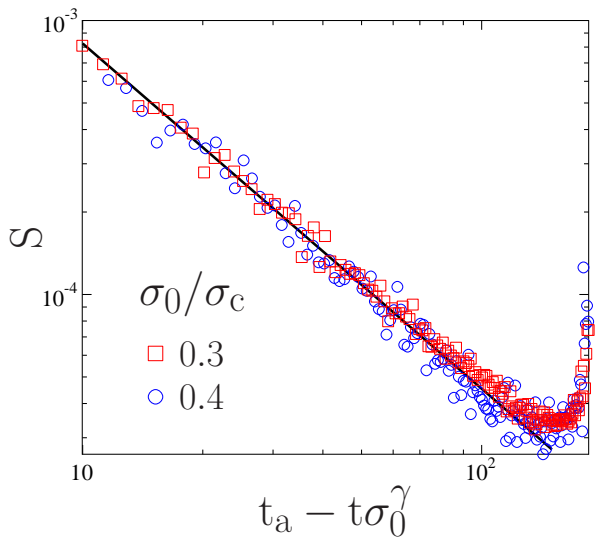,
  width=8.0cm}
   \caption{{\it (Color online)} 
Scaling plot of deformation-time curves measured experimentally on
asphalt
specimens \cite{kun_asphalt_jstat_2007}. The scaling function $S$,
obtained by rescaling the two axis, has a power law dependence on the
time-to-failure. 
}
\label{fig:deform}
  \end{center}
\vspace*{-0.55cm}
\end{figure}
On the macrolevel the process of fatigue is characterized by
the evolution of deformation $\varepsilon(t)$ of the specimen, which
is related to $p(t)$ as $p(t)=E\varepsilon(t)$, where $E=1$ is the
Young modulus of fibers. Neglecting immediate breaking and healing,
Eq.\ (\ref{eq:eom}) can be transformed into a differential equation
for the number $N_b$ of broken fibers as $dN_b/dt=
af(c(t))p^{\gamma}N$, where $f(c)=dF/dc$. Using $p(t)=N\sigma_0/
\left(N-N_b(t)\right)$, for uniformly distributed threshold values the
exact solution of the equation of motion Eq.\ (\ref{eq:eom}) reads 
\beq{
\varepsilon(t) = \sigma_0 \left[(t_f-t)/t_f\right]^{-1/(1+\gamma)} \\
 \mbox{~and~} \\ t_f = \frac{\sigma_0^{-\gamma}}{a(1+\gamma)},
\label{eq:basq_analit}
} 
where $t_f$ denotes the lifetime of the system.
Equation~(\ref{eq:basq_analit}) shows that damage accumulation leads to a
finite time singularity where the deformation $\varepsilon(t)$ of the
system has a power law divergence with an exponent determined by
$\gamma$. It is important to emphasize that $t_f$ has a power law
dependence on the external load $\sigma_0$ in agreement with Basquin's
law of fatigue found experimentally in a broad class of materials
\cite{basquin_1910,krajcinovic_damagebook_1996,krajcinovic_1}. The
Basquin exponent of the model therefore coincides with that of 
the microscopic degradation law $\alpha = \gamma$.
Another interesting outcome of the derivation is that
the macroscopic deformation $\varepsilon(t)$ of a specimen undergoing
fatigue fracture obeys the generic scaling form
$\varepsilon(t) = \sigma_0^{\delta}S(t\sigma_0^{\beta})$,
where the scaling function $S$ has the property $S(t\sigma_0^{\beta}) \sim
(t_a - t\sigma_0^{\beta})^{-1/(1+\gamma)}$, with $t_a =a(1+\gamma)$ 
and the scaling exponents are $\delta=1$ and $\beta = \gamma$. 
Figure~\ref{fig:deform} presents a verification of this scaling law on
experimental 
data from asphalt specimens obtained at two different load values
\cite{kun_asphalt_jstat_2007}. The good quality
data collapse obtained by rescaling the two axis and the power law
behavior of $S$ as a function of the time-to-failure demonstrates the
validity of our scaling relation.

\begin{figure}%[!h]
  \begin{center}
\psfrag{aa}{\large $a)$}
\psfrag{bb}{\large $b)$}
\epsfig{bbllx=45,bblly=470,bburx=340,bbury=750,file=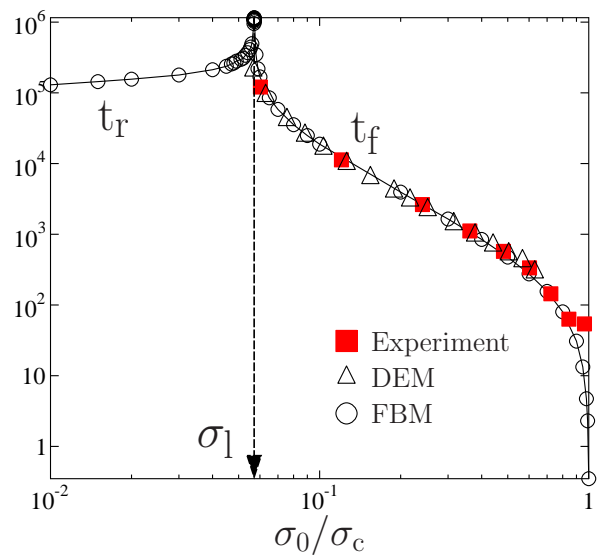,
  width=7.6cm} \caption{{\it (Color online)} Characteristic time
  scales $t_r$ and $t_f$ of the system. The complete
  FBM corresponding 
  to Eq.\ (\ref{eq:eom}) which includes immediate breaking and healing
  is solved numerically. For $\sigma_0 > \sigma_l$ we see Basquin's
  law and both models provide a very good fit of the lifetime data of
  asphalt.  The fatigue limit $\sigma_l$ is indicated by the vertical dashed
  line.  }
\label{fig:phase_diag}
  \end{center}
\vspace*{-0.6cm}
\end{figure}
\begin{figure}%[!h]
  \begin{center}
\epsfig{bbllx=35,bblly=460,bburx=340,bbury=750,file=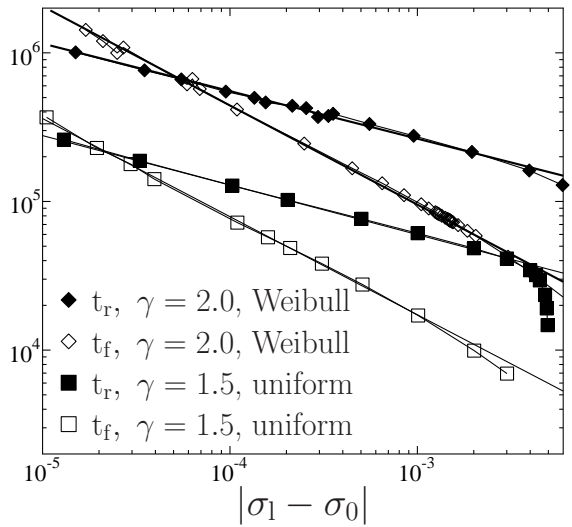,
  width=7.6cm} \caption{The relaxation time $t_r$ and lifetime $t_f$
  as a function of $|\sigma_l-\sigma_0|$ 
  for different disorder distributions (uniform and Weibull) and
  $\gamma$ exponents. The straight lines have slopes $-1/3$ and
  $-2/3$.}
\label{fig:critical}
  \end{center}
\vspace*{-0.6cm}
\end{figure}

Healing dominates if for a fixed load $\sigma_0$ the memory time
$\tau$ is smaller than the lifetime obtained without healing $\tau
\lesssim t_f(\sigma_0,\tau=+\infty)$. Then, a threshold load
$\sigma_l$ emerges below which the system relaxes, {\it i.e.}, the
deformation $\varepsilon(t)$ converges to a limit value with a
characteristic relaxation time $t_r$ resulting in an infinite
lifetime.  Figure~\ref{fig:phase_diag} presents the characteristic
time scale of the system varying the external load over a broad
range. The results from numerical simulations with the complete FBM
(i.e., including immediate breaking and healing) are in excellent
agreement with the measured lifetime of asphalt samples for $\sigma_0
> \sigma_l$, recovering also the Basquin exponent
\cite{kun_asphalt_jstat_2007}. 
The regime below $\sigma_l$ is of particular importance in geodynamics
where memory effects take place during cyclic loading of rocks with a
stress amplitude increasing from one cycle to the next
\cite{memory_effect}. 
%Interestingly, the regime below 
%$\sigma_l$ has not yet been studied experimentally. 
It is important to
note that approaching the fatigue limit $\sigma_l$ from either side,
the characteristic time scale diverges. Figure 
\ref{fig:critical} shows that both the relaxation time $t_r$ and the
lifetime $t_f$ follow a power law as a function of the difference from
the fatigue limit with distinct exponents: $t_r \sim
(\sigma_l-\sigma_0)^{-1/3}$ and $t_f\sim
(\sigma_0-\sigma_l)^{-2/3}$. We stress that the exponents neither
depend on the disorder distributions ($F$ and $G$) nor on the details
of the damage law ($a$, $\gamma$ and $\tau$), {\it i.e.}, they are
universal implying a continuous phase transition at the fatigue limit
$\sigma_l$ between partial failure and macroscopic fracture (see
Fig.\ \ref{fig:critical}).

Our calculations revealed that the Basquin law of lifetime emerges on
the macrolevel as a consequence of the competition between the two
microscopic failure mechanisms of fibers. Rewriting Eq.\
(\ref{eq:eom}) in the form of the constitutive equation of simple FBMs
as $\sigma_0/\left[1-F(c(t))\right] = \left[1-G(p(t))\right]p(t)$ it
can be seen that the slow damage process on the left hand side
quasi-statically increases the load on the system: ageing fibers
accumulate damage and break slowly one-by-one in the increasing order
of their damage thresholds $c_{th}^i$. After a number $\Delta_d$ of
damage breakings, the emerging load increment on the remaining intact
fibers can trigger a burst of immediate breakings. Since load
redistribution and immediate breaking occur on a much shorter time
scale than damage accumulation, the entire fatigue process can be
viewed on the microlevel as a sequence of bursts of immediate
breakings triggered by a series of damage events happening during
waiting times $T$, {\it i.e.}, the time intervals between the bursts.
%\begin{figure}%[!h]
%  \begin{center}
%\epsfig{bbllx=20,bblly=465,bburx=345,bbury=750,file=damlav_paper.eps,
%  width=7.6cm} \caption{{\it (Color online)} Size distribution of
%  damage sequences $P(\Delta_d)$ obtained from numerical simulations
%  with the FBM. Good quality data collapse is obtained by rescaling
%  the two axis by $\sigma_0$, in agreement with the analytic
%  predictions. Inset: DEM simulations in two dimensions give the same
%  qualitative behavior but with a higher exponent. }
%\label{fig:dam_sequence}
%  \end{center}
%\vspace*{-0.6cm}
%\end{figure}
The microscopic failure process is characterized by the size
distribution of bursts $P(\Delta)$, damage sequences $P(\Delta_d)$,
and by the distribution of waiting times $P(T)$. At small loads
$\sigma_0\ll\sigma_c$ most of the fibers break in long damage
sequences, because the resulting load increments do not suffice to
trigger bursts. 
%In this case the system behaves similar to a simple
%FBM under quasi-static loading where the loading process was stopped
%much below $\sigma_c$. 
Consequently, the burst size distribution
$P(\Delta)$ has a rapid exponential decay. Increasing $\sigma_0$ the
total number of 
bursts $n_b$ increases linearly $n_b \sim \sigma_0$ and a power law
regime of burst sizes emerges $P(\Delta) \sim \Delta^{-\xi}$ with the
well-known mean field exponent of FBM $\xi=5/2$
\cite{hansen_crossover_prl}. When macroscopic failure is approached
$\sigma_0\to\sigma_c$ the failure process accelerates such that the
size $\Delta_d$ and duration $T$ of damage sequences decrease, while
they trigger bursts of larger sizes $\Delta$, and finally macroscopic
failure occurs as a catastrophic burst of immediate failures. Since in
the limiting case of $\sigma_0 \to \sigma_c$ a large number of weak
fibers breaks in the initial burst, we found that the distribution
$P(\Delta)$ has a 
crossover to a smaller exponent $\xi=3/2$, in agreement with Ref.\
\cite{hansen_crossover_prl}. After the linear increase, the number of
bursts $n_b$ has a maximum at $\sigma_0/\sigma_c\approx 0.4$ and
rapidly decreases to 1 as $\sigma_c$ is approached. All these results
are independent of $\gamma$, $a$, and $\tau$.

\begin{figure}%[!h]
  \begin{center}
\epsfig{bbllx=15,bblly=465,bburx=345,bbury=750,file=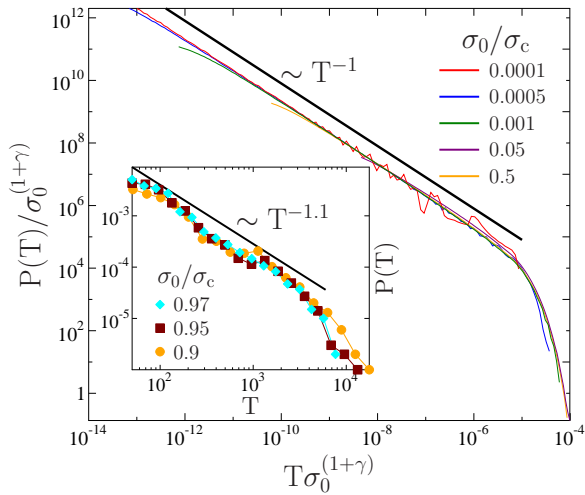,
  width=7.6cm} \caption{{\it (Color online)} Rescaled plot of
  waiting time distributions $P(T)$ obtained by FBM simulations.
  Inset: Corresponding results of two-dimensional DEM simulations.}
\label{fig:waiting}
  \end{center}
\vspace{-0.5cm}
\end{figure}

Since damage events increase the load on the remaining
intact fibers until an immediate breaking is triggered, the size of
damage sequences $\Delta_d$ is independent of the damage
characteristics $c(t)$ and $F(c_{th})$ of the material, instead, it is
determined by the load bearing strength distribution $G(p_{th})$ of
fibers. 
%Whilst $\Delta_d$ is the number of
%fibers which can be removed from the bundle without giving rise to an
%immediate breaking, the size distribution of damage sequences
%$P(\Delta_d)$ can be obtained analytically as an integral
%$P(\Delta_d)\sim\int_{p_0}^{p_c}e^{-\Delta_dpg(p)/[1-G(p)]}dp$, where
%$g(p)=dG/dp$, and load dependence arises only through 
%$p_0=p_0(\sigma_0)$. 
Under broad conditions this mechanism
leads to an universal power law form with an exponential cutoff
$P(\Delta_d) \sim
\Delta_d^{-1}\exp{(-\Delta_d/\left<\Delta_d\right>)}$, where
$\left<\Delta_d\right> \sim \sigma_0^{-1}$.
The damage law $c(t)$ of the material controls the time scale of the
process of fatigue fracture through the temporal sequence 
of single damage events. 
In damage sequences fibers break in the increasing order of their
damage thresholds $c_{th}^i$ which determine the time intervals
$\Delta t$ between consecutive fiber breakings. 
%Hence, the distribution of inter-event times $P(\Delta t)$ can be
%obtained analytically based on the property that the probability
%distribution of the $i$th largest element of a sorted sequence of $N$
%thresholds is sharply peaked for each $i$ for large enough $N$
%values. 
Analytic calculations showed that $P(\Delta t)$ has an explicit
dependence on $\gamma$ as $P(\Delta t) \sim 
\Delta t^{-(1-1/\gamma)}$, however, the duration of sequences
$T=\sum_{j=1}^{\Delta_d}\Delta t_{j}$, {\it i.e.}, the waiting times
between bursts follow an universal power law distribution $P(T)\sim
T^{-1}\exp{(-T/\left<T\right>)}$, where only the cutoff has
$\gamma$-dependence $\left<T\right>\sim\sigma_0^{-(1+\gamma)}$ (see
Fig.\ \ref{fig:waiting}). 

The macroscopic lifetime $t_f$ of a finite system can be related to
characteristic quantities of the microscopic failure process as $t_f =
\sum_{i=1}^{n_b}T_i$, from which the average lifetime can be
obtained in the form $t_f \approx \left<n_b\right> \left<T\right>$. In
the load regime where the generic scaling laws of the distributions
$P(\Delta)$, $P(\Delta_d)$, and $P(T)$ prevail, this leads to the form
$t_f \sim \sigma_0^{-\gamma}$ in agreement with the Basquin law Eq.\
(\ref{eq:basq_analit}) of the system. The results demonstrate that the
Basquin law of lifetime on the macro-scale is a fingerprint of the
scale-free microscopic bursting activity, with the material dependence
entering only through the damage law determining the waiting times
between bursts. Experimentally, the microscopic fracture process
underlying fatigue can be monitored by the acoustic emission technique
and by direct optical observations
\cite{zapperi_alava_statmodfrac,santucci_prl_subcrit_2004,krajcinovic_damagebook_1996,krajcinovic_1}.
Sub-critical cracking has recently been found to produce a power law
distribution of step sizes of the advancing crack in agreement
with our predictions on the size distribution of bursts
\cite{santucci_prl_subcrit_2004}. 

In order to study the effect of stress concentration and
crack growth in fatigue fracture, we also developed a discrete element
model (DEM) \cite{humberto} in which we discretize a two-dimensional 
disc-shaped specimen in terms of randomly shaped convex polygons connected 
by elastic beams. 
The beams fail either due to immediate breaking or
damage which are coupled in a single failure variable
$q(t)=p(t)+a\int_0^{t}e^{-(t-t')/\tau}p(t')^{\gamma}dt'$.
Here $p(t)$ describes the deformation state of the beam
taking into account both stretching and bending $p(t) = (\varepsilon/\varepsilon_{th})^2+
\max{(|\Theta_1|,|\Theta_2|)}/\Theta_{th}$, being $\varepsilon$ the
longitudinal deformation, $\Theta_1$ and $\Theta_2$ the bending angles 
at the two ends of the beam, and $\varepsilon_{th}$ 
and $\Theta_{th}$ denote the threshold values a beam can sustain 
under stretching and bending, respectively. As a
consequence, the parameters $a$, $\gamma$, and $\tau$ play the same role as their
counterparts in our FBM. The time evolution of the system is followed by numerically
solving the equations of motion of polygons. The breaking
criterion $q(t)>1$ is evaluated at each time step and beams
which fulfil the condition are removed \cite{humberto}. We
study the fatigue fracture under diametric compression of discs
with constant stress $\sigma_0$ (Brazil test). Figure~\ref{fig:phase_diag} 
shows that DEM provides also an excellent fit of the lifetime data 
of asphalt specimens \cite{kun_asphalt_jstat_2007}.
DEM simulations revealed that in the presence of stress concentrations
bursts are spatially correlated and they can be identified as sudden
advancements of slowly growing cracks. DEM results on burst
characteristics also show
power law behavior as the mean field FBM, but with different exponents
due to the two-dimensionality of the model. The localized
stress concentration built up around cracks gives rise to higher
values of the exponents of the size distribution of bursts $P(\Delta)
\sim \Delta^{-2.7}$, and of damage sequences
$P(\Delta_d)\sim \Delta_d^{-1.8}$,
while for the waiting time distribution $P(T)$ the DEM exponent falls
very close to the mean field value (see Fig.\ \ref{fig:waiting}). The
results proved to be independent of the value of $\gamma$.

Although the exponent of Basquin's law depends on the microscopic
damage accumulation, we found an astonishing spectrum of universal
features hidden behind this originally empirical law. On one hand we
discovered in the experimentally relevant situation of finite damage
memory a continuous phase transition between partial failure and
macroscopic rupture. On the microscopic level of individual breaking
events we showed that the separation of time scales of the two
competing failure mechanisms leads to a bursting activity, where we
disclosed several universal scaling laws in the distributions and
determined their exponents as well in mean-field as in two dimensions.
In summary our approach provides a direct connection between the
microscopic mechanisms constituting the main ingredients of the model
(i.e., immediate breaking, damage accumulation and healing of
microcracks) and the macroscopic behavior of the fatigue process. 
The (macroscopic)
exponent from Basquin's law coincides with the (microscopic) exponent
of the degradation law, namely $\alpha=\gamma$. Following a slightly
different pathway, our methodology is also capable to show explicitly
the bridge between the (universal) mechanism related with the
scale-free bursting activity at the micro-scale and the
(non-universal) lifetime law of the material at the macro-scale. 

This work opens up new experimental challenges. 
Our scaling relation of the macroscopic deformation
should be verified on various types of materials, after which it could
help to extract the relevant information from fatigue life
measurements. For instance, it
would be interesting to check our theoretical predictions with fatigue
measurements performed at very low external loads, i.e., for $\sigma_0
\approx \sigma_l$. More precisely, in the infinite lifetime limit,
$\sigma_0 \lesssim \sigma_l$, the experimental confirmation of the
power-law variability with load of the relaxation time should
certainly provide some considerable insight on the role of healing in
the entire fatigue process. For similar reasons, it would be also
interesting to verify the distinct lifetime behavior obtained from the
model in the other limit of low external loads, $\sigma_0 \gtrsim
\sigma_l$. Finally, another interesting outcome from our study is the 
statistical behavior related with the bursting activity during the
fatigue process of a given material. According to our analysis, both
the size $\Delta_d$ of damage sequences and magnitude $T$ of waiting
times between bursts should obey universal power-law distributions
that might reflect the intrinsic features of the typical restructuring
events taking place at the microscopic level. 
As a possible monitoring technique, acoustic
emission measurements could be conducted in 
conjunction with fatigue experiments to confirm our claim for
universality behind Basquin's law.

We thank the Brazilian agencies CNPq, CAPES, FUNCAP and FINEP, and the
Max Planck prize for financial support. F.\ Kun was supported by
OTKA T049209.


\begin{thebibliography}{100}

\bibitem{zapperi_alava_statmodfrac}
M.\ Alava, P.\ K.\ Nukala, and S.\ Zapperi, Adv.\ Phys.\ {\bf 55}, 349
(2005).

\bibitem{schwarz_adhesionbio_prl_2004}
T.\ Erdmann and U.\ Schwarz, Phys.\ Rev.\ Lett.\ {\bf 92}, 108102
(2004).

\bibitem{santucci_prl_subcrit_2004} 
S.\ Santucci et al., Phys.\ Rev.\ Lett.\ {\bf 93}, 095505 (2004).

\bibitem{sornette_thermal_prl_2005}
D.\ Sornette and G.\ Ouillon, Phys.\ Rev.\ Lett.\ {\bf 94}, 038501
(2005).

\bibitem{nechad_sornette_prl_2005}
H.\ Nechad et al., Phys.\ Rev.\ Lett.\ {\bf 94}, 045501 (2005);R.\ C.\
Hidalgo, F.\ Kun, and H.\ J.\ Herrmann, Phys.\ Rev.\ E {\bf 65},
032502 (2002).

\bibitem{sornette_fusefatigue_prl1992}
D.\ Sornette and C.\ Vanneste, Phys.\ Rev.\ Lett.\ {\bf 68}, 612
(1992); H.\ J.\ Herrmann, J.\ Kert\'esz, and L.\ de Arcangelis,
Europhys.\ Lett.\ {\bf 10}, 147 (1989).

\bibitem{farkas_crystal_fatigue_prl_2005}
D.\ Farkas, M.\ Willemann, and B.\ Hyde, Phys.\ Rev.\ Lett.\ {\bf 94},
165502 (2005); D.\ Sornette, T.\ Magnin, and Y.\ Brechet, Europhys.\
Lett.\ {\bf 20}, 433 (1992).

\bibitem{marone_nature_healing_1998}
C.\ Marone, Nature {\bf 391}, 69 (1998).

\bibitem{basquin_1910}
O.\ H.\ Basquin, Proc.\ ASTM {\bf 10}, 625 (1910).

\bibitem{krajcinovic_damagebook_1996}
D.\ Krajcinovic, {\it Damage Mechanics}, (Elsevier, Amsterdam, 1996);
S.\ Suresh, {\it Fatigue of Materials}, (Cambridge University Press, 2006).

\bibitem{krajcinovic_1}A.\ Rinaldi et al., Int.\ J.\ Fatigue {\bf 28},
1069 (2006); M.\ E.\ Biancolini et al., ibid,
1820.

\bibitem{sornette_prl_78_2140}
J.\ V.\ Andersen, D.\ Sornette, and K.\ Leung, Phys.\ Rev.\ Lett.\
{\bf 78}, 2140 (1997); F.\ Kun, S.\ Zapperi, and H.\ J.\ Herrmann,
Eur.\ Phys.\ J.\ B {\bf 17}, 269 (2000); R.\ C.\ Hidalgo et al.,
Phys.\ Rev.\ Lett.\ {\bf 89}, 205501 (2002).

\bibitem{humberto}
H.\ A.\ Carmona et al., Phys.\ Rev.\ E {\bf 75}, 046115 (2007).

\bibitem{kun_asphalt_jstat_2007}
F.\ Kun et al., J.\ Stat.\ Mech.\ P02003 (2007); M.\ J. Alava, J.\
Stat.\ Mech.\ N04001 (2007).

\bibitem{hansen_crossover_prl}
M.\ Kloster, A.\ Hansen, and P.\ Hemmer, Phys.\ Rev.\ E {\bf 56}, 2615
(1997); S.\ Pradhan, A.\ Hansen, and P.\ C.\ Hemmer, Phys.\ Rev.\ Lett.\ {\bf
95}, 125501 (2005).
   
\bibitem{memory_effect} A.\ Lavrov, Int.\ J. Rock Mech.\ Min.\ Sci.\
{\bf 40}, 151 (2003).
\end{thebibliography}
\end{document}